\documentclass[12pt,paper,english,preprint,aps,showkeys,showpacs]{revtex4-1}
\usepackage{amsmath,amsthm,amsfonts,amssymb}
\usepackage{amsthm,txfonts}
\usepackage{mathrsfs}
\usepackage{braket}
\usepackage{hyperref}
\usepackage{color}
\usepackage{hyperref}
\usepackage{slashed}

\newcommand{\ie}{\begin{equation}}
\newcommand{\fe}{\end{equation}}
\newcommand{\se}{\begin{eqnarray}}
\newcommand{\ff}{\end{eqnarray}}

\begin{document}

\title{Effects of spontaneous Lorentz violation in gravitational waves solutions}


\href{<https://workshopparticulas.wixsite.com/ufcg>}{\small{\bf WPF 2018}, Workshop on Particles and Fields - November 28 - 30, 2018, Campina Grande - PB, Brazil.}

\author{Kevin M. Amarilo}
\email{kevin@fisica.ufc.br}
\affiliation{Universidade Federal do Cear\'a (UFC), Departamento de F\'isica,\\ Campus do Pici, Fortaleza - CE, C.P. 6030, 60455-760 - Brazil.}


\author{Mapse Barroso F. Filho}
\email{mapse@fisica.ufc.br}
\affiliation{Universidade Federal do Cear\'a (UFC), Departamento de F\'isica,\\ Campus do Pici, Fortaleza - CE, C.P. 6030, 60455-760 - Brazil.}

\author{Roberto V. Maluf}
\email{r.v.maluf@fisica.ufc.br}
\affiliation{Universidade Federal do Cear\'a (UFC), Departamento de F\'isica,\\ Campus do Pici, Fortaleza - CE, C.P. 6030, 60455-760 - Brazil.}


\date{\today}

\begin{abstract}

Gravitational wave solutions with spontaneous Lorentz symmetry breaking are studied. This breaking is triggered by a vector field, known as the bumblebee field. A brief analysis of the main consequences of this symmetry breaking on free-waves solutions and in the presence of matter sources are presented. For the free case, the modified wave equation for the gravitational perturbation field is solved, and we compare the polarization states with the usual solution. Considering a current that acts as a source of the radiation, we solve the equation of motion using the Green function method. Finally, considering that the source has slow motion, we derive the new quadrupole formula.
\end{abstract}


\maketitle

\section{Introduction}

Gravitational waves detection was one of the most important events for the high energy physics of the XXI century. On February 11th, 2016 the laboratory VIRGO and LIGO announced that the first detection was made on September 14th, 2015, in an event called GW150914 \cite{abbott1}. The second detection was done in December 26th, 2015, this event is called by GW151226. The detected waves of both events are due to coalescence of two black holes located at distances of approximately 410 Mpc and 440 Mpc \cite{abbott2}. 

On the other hand, the construction of a quantum theory of gravity has been a great challenge to many theoretical physicists for many years. The problem in the development of such theory is that its effects are expected to be relevant only in energy scales of the Planck mass $m_P \simeq 1.22 \times 10^{19}$ GeV, therefore, we lack experimental evidence, and the construction of the theory relies only on theoretical consistency.

Although the Planck scale is still inaccessible, many models beyond the general relativity (GR) and the standard model (SM) lead to deviations from the expected predictions at low energy in specific regimes. One of these phenomena is the violation of Lorentz symmetry. The investigation of the Lorentz symmetry breaking (LSB) is addressed to an effective field theory known as standard model extension (SME), in which the usual fields of the SM are coupled with fixed background fields that trigger the Lorentz and CPT breaking \cite{SME1}.

Among the fields that break the Lorentz symmetry, the bumblebee field $B_\mu$ is one relatively simple \cite{KS}. The symmetry is spontaneously broken by the dynamics of $B_\mu$ that acquires a nonzero vacuum expectation value (VEV) \cite{Bluhm2005, BluhmFung}. This model was first considered in the context of string theory with the LSB being triggered by the potential $V(B^\mu) = \lambda(B^\mu B_\mu \mp b^2)^2/2$.

In this work, we analyze the modifications due to the LSB in the polarization and production of gravitational waves (GW) in the weak field regime. We start investigating the consequences of the spontaneous breaking of Lorentz symmetry, triggered by the bumblebee vector field. The modified wave equation for the perturbation field is solved, and we compare the polarization tensor of the modified gravitational wave solution with the usual case. Next, a current $J_{\mu\nu}$ is added to the model and a Green function is derived for the timelike $b^\mu = (b^0, 0)$ and spacelike $b^\mu = (0, {\bf b})$ configurations for the bumblebee VEV. In sequence, we obtain a modified quadrupole formula for the graviton. With these results, the perturbation for the modified equation is compared with the usual one, showing the modifications in the theory.


\section{Modifying the Graviton Wave Equation}

The minimal extension of the gravity theory including the Lorentz-violating terms is given by the action \cite{KosteleckyG1}
\begin{equation}
    S = S_{EH} + S_{LV} + S_{matter}.
\end{equation}
The first term refers to the usual Einstein-Hilbert action,
\begin{equation}
    S_{EH} = \int d^4x \; \sqrt{-g} \frac{2}{\kappa^2} (R - 2\Lambda),
\end{equation}where $R$ is the curvature scalar and $\Lambda$ is the cosmological constant, which will not be considered in this analysis. The $S_{LV}$ term consists in the coupling between the bumblebee field and the curvature of spacetime. The leading terms are
\begin{equation}\label{eq:SLV}
    S_{LV} =  \int d^4x \; \sqrt{-g} \frac{2}{\kappa^2} (uR + s^{\mu\nu}R_{\mu\nu} + t^{\alpha\beta\mu\nu} R_{\alpha\beta\mu\nu}),
\end{equation}where $u$, $s^{\mu\nu}$ and $t^{\alpha \beta \mu \nu}$ are dynamical fields with zero mass dimension. And the last term takes account the matter-gravity couplings, which in principle should include all fields of the standard mode model as well as the possible interactions with the coefficients $u$, $s^{\mu \nu}$ and $t^{\alpha \beta \mu \nu}$.

The dynamics of the bumblebee field $B^\mu$ is dictated by the action \cite{MalufGravity1}
\begin{equation}
    S_B = \int d^4x \; \sqrt{-g} \left[ -\frac{1}{4} B^{\mu \nu} B_{\mu \nu} + \frac{2 \xi}{\kappa^2} B^\mu B^\nu R_{\mu\nu} - V(B^\mu B_\mu \mp b^2) \right],
\end{equation}where we introduced the field strength $B_{\mu \nu} = \partial_\mu B_\nu - \partial_\nu B_\mu$ in analogy with the electromagnetic field tensor $F_{\mu\nu}$. In fact, the bumblebee models are not used only as a toy model to investigate the excitation originated from the LSB mechanism, but also as an alternative to $U(1)$ gauge theory for the photon \cite{escobar}. In this theory, the photon appears not as a fundamental particle, but as a Nambu-Goldstone mode due to spontaneous Lorentz violation \cite{Hernaski:2014jsa}.

The gravity-bumblebee coupling can be represented by the action (\ref{eq:SLV}) defining
\begin{equation}
    u = \frac{1}{4}\xi B^\alpha B_\alpha, \quad s^{\mu\nu} = \xi \left(B^\mu B^\nu -\frac{1}{4} g^{\mu\nu} B^\alpha B_\alpha \right), \quad t^{\alpha \beta \mu \nu} = 0.
\end{equation}

The potential $V(X)$ is defined as
\begin{equation}
V=\frac{\lambda }{2}\left( B^{\mu }B_{\mu }\mp b^{2}\right) ^{2},
\end{equation}and it is responsible for triggering the spontaneous breakdown of diffeomorphism and Lorentz symmetry. Here $b^2$ is a positive constant that stands for the nonzero expectation value of this field.

In the pursuit of the influence of the gravity-bumblebee coupling on the graviton, we consider the linearized version of the metric $g_{\mu \nu}$ with the Minkowski background and the bumblebee field $B_\mu$, which is split into the vacuum expectation values $b_\mu$ and the quantum fluctuations $\Tilde{B}^\mu$
\begin{eqnarray}
g_{\mu\nu} & = & \eta_{\mu\nu}+\kappa h_{\mu\nu},  \notag \\
B_{\mu} & = & b_{\mu}+\tilde{B}_{\mu},  \notag \\
B^{\mu} & = & b^{\mu}+\tilde{B}^{\mu}-\kappa b_{\nu}h^{\mu\nu}.
\label{eq:Expansion1}
\end{eqnarray}

Following the procedure described in Refs. \cite{MalufGravity1,Bailey:2006slv}, the solution for the linearized bumblebee equation of motion in the momentum space is
\begin{equation}
\tilde{B}^{\mu }=\frac{\kappa p^{\mu }b_{\alpha }b_{\beta }h^{\alpha \beta}
}{2b\cdot p}+\frac{2\sigma b_{\alpha }R^{\alpha \mu }}{p^{2}}-\frac{2\sigma
p^{\mu }b_{\alpha }b_{\beta }R^{\alpha \beta }}{p^{2}b\cdot p}+\frac{\sigma
p^{\mu }R}{4\lambda b\cdot p}-\frac{\sigma b^{\mu }R}{p^{2}}+\frac{\sigma
p^{\mu }b^{2}R}{p^{2}b\cdot p},
\end{equation}
with $p^\mu = (p^0, {\bf p}), \sigma = (2\xi / \kappa^2 )$, while $R_{\mu\nu}$ and $R$ are taken in their linearized form. This solution can be inserted in the Lagrangian (\ref{eq:SLV}) as specified in \cite{MalufGravity1} leading to the following effective Lagrangian density  
\begin{eqnarray}
\mathcal{L}_{\mbox{LV}} &=&\xi \left[ p^{2}b_{\mu }b_{\nu }h^{\mu \nu }h_{\
\alpha }^{\alpha }+\frac{1}{2}\left( b\cdot p\right) ^{2}\left( h_{\ \alpha
}^{\alpha }\right) ^{2}\right.   \notag \\
&-&\left. \frac{1}{2}\left( b\cdot p\right) ^{2}h^{\mu \nu }h_{\mu \nu
}+p^{2}b_{\mu }b_{\nu }h^{\mu \alpha }h_{\ \alpha }^{\nu }-\left( b_{\mu
}b_{\nu }p_{\alpha }p_{\beta }+b_{(\mu }p_{\nu )}b_{(\alpha }p_{\beta
)}\right) h^{\mu \nu }h^{\alpha \beta }\right]   \notag \\
&+&\frac{4\xi ^{2}}{\kappa ^{2}}\left[ \left( -2p^{2}b_{\mu }b_{\nu
}-2b^{2}p_{\mu }p_{\nu }+4b\cdot pb_{(\mu }p_{\nu )}-\frac{p^{2}p_{\mu
}p_{\nu }}{4\lambda }\right) h^{\mu \nu }h_{\ \alpha }^{\alpha }\right.
\notag \\
&+&\left( 2b_{\mu }b_{\nu }p_{\alpha }p_{\beta }-b_{(\mu }p_{\nu
)}b_{(\alpha }p_{\beta )}+\frac{b^{2}p_{\mu }p_{\nu }p_{\alpha }p_{\beta }}{%
p^{2}}-\frac{2b\cdot pp_{\mu }p_{\nu }b_{(\alpha }p_{\beta )}}{p^{2}}+\frac{%
p_{\mu }p_{\nu }p_{\alpha }p_{\beta }}{4\lambda }\right) h^{\mu \nu
}h^{\alpha \beta }  \notag \\
&+&\left. \left( b^{2}p^{2}-\left( b\cdot p\right) ^{2}+\frac{p^{4}}{%
4\lambda }\right) \left( h_{\ \alpha }^{\alpha }\right) ^{2}+\left(
p^{2}b_{\mu }b_{\nu }-2b\cdot pb_{(\mu }p_{\nu )}+\frac{\left( b\cdot
p\right) ^{2}p_{\mu }p_{\nu }}{p^{2}}\right) h^{\mu \lambda }h_{\ \lambda
}^{\nu }\right] +\mathcal{O}(h^{3}),  \notag \\
&&
\label{lvf}
\end{eqnarray} where $b\cdot p = b_{\mu}p^{\mu}$ with $p^{\mu}=(p^{0},{\bold p})$ and $b^{\mu}=(b^{0},{\bold b})$. It is worth noting that this Lagrangian is not invariant under the usual gauge transformation of the metric perturbation $(h_{\mu\nu} \rightarrow h_{\mu\nu} + i p_\mu \zeta_\nu + i p_\nu \zeta_\mu)$. Hence, the effect of the bumblebee field on the graviton dynamics is encoded in the effective Lagrangian density
\begin{equation}\label{effectivel}
    \mathcal{L}_{kin} = \mathcal{L}_{EH} +\mathcal{L}_{gf}+ \mathcal{L}_{LV},
\end{equation}where $\mathcal{L}_{EH}$ is the usual Einstein-Hilbert Lagrangian and, $\mathcal{L}_{gf}$ is one convenient gauge fixing term \cite{MalufGravity1}.

The equation of motion to the graviton field obtained from the effective Lagrangian \eqref{effectivel} is
\begin{equation}\label{eq:motiongravmod}
    \Hat{\mathcal{O}}_{\mu\nu,\alpha\beta} h^{\alpha\beta} = 0,
\end{equation}with the operator $\Hat{\mathcal{O}}$ defined by
\begin{equation}
    \mathcal{L}_{kin} = -\frac{1}{2} h^{\mu\nu} \Hat{\mathcal{O}}_{\mu\nu,\alpha \beta} h^{\alpha \beta},
\end{equation}such that it is symmetric in the indices $(\mu\nu)$, $(\alpha \beta)$ (the same way as $h^{\mu\nu}$ and $h^{\alpha \beta}$) and under the interchange of the pairs $(\mu\nu)$ and $(\alpha \beta)$.

The Eq. (\ref{eq:motiongravmod}) is then saturated with $p^\mu p^\nu$, $b^\mu b^\nu$, $p^{(\mu} b^{\nu)}$ and the Minkowski metric, revealing the following constraints \cite{Maluf2014}
\begin{equation}\label{eq:const1}
    p_\mu p_\nu h^{\mu\nu} = 0, \quad b_\mu b_\nu h^{\mu\nu} = 0, \quad p_{(\mu} b_{\nu)} h^{\mu\nu} = 0, \quad h = 0.
\end{equation}

Even more constraints can be found saturating (\ref{eq:motiongravmod}) with $p^\mu$ and $b^\mu$ alone, which leads to
\begin{equation}\label{eq:const6}
    p_\mu h^\mu_\nu = 0, \quad b_\mu h^\mu_\nu = 0.
\end{equation}

The Eqs. (\ref{eq:const1})-(\ref{eq:const6}) consist of a set of 12 constraints that can reduce the 14 degrees of freedom of the theory (ten of the graviton and four of the bumblebee field). Therefore, we are left with two physical degrees of freedom, which is the same number as the usual Einstein-Hilbert graviton \cite{Maluf2014}.

We apply the Eqs. (\ref{eq:const1})-(\ref{eq:const6}) to the Eq. (\ref{eq:motiongravmod}), and it simplifies to
\begin{equation}\label{hmneq}
    [p^2 + \xi (b \cdot p)^2]h_{\mu\nu} = 0,
\end{equation}that is the correct energy-momentum dispersion relation associated with the physical pole \cite{Maluf2014}. Therefore, we can conclude that the spontaneous Lorentz violation triggered by the bumblebee field modified the Einstein-Hilbert dispersion relation as $p^2 = 0 \rightarrow p^2 + \xi (b \cdot p)^2 = 0$. Moreover, the nonminimal coupling $B^\mu B^\nu R_{\mu\nu} $ has not produced massive modes for the graviton.


\section{Solutions of the modified graviton equation}

From Eq. \eqref{hmneq}, the dispersion relation  associated with the graviton field is  
\begin{equation}
    \label{dr}
    p_0 = \left|{\bf p}\right|\left[\frac{\xi b_0\left|{\bf b}\right|\cos\Psi \pm \sqrt{1 + \xi b_0^2 - \xi \left|{\bf b}\right|^2 \cos^2\Psi}}{1+\xi b_0^2}\right],
\end{equation}where $(b\cdot p)=b_{0}p_{0}-\left|{\bf p}\right|\left|{\bf b}\right|\cos\Psi$. Real roots are assured by the condition $b_{0}^{2}>\left|{\bf b}\right|^{2}\cos^2\Psi$. We can note that there are two solutions in the dispersion relation \eqref{dr}) which have the generic form $p_{0\pm} = f({\bf p}) \pm \sqrt{g({\bf p})}$. However, these solutions are not independent since the expression \ref{hmneq} is invariant under the substitution $p_\mu \rightarrow -p_\mu$. Thus, only one of them will be considered, so that the solution used hereafter will be $p_{0}=p_{0+}$ \cite{Hernaski:2014jsa}.

The solution of the graviton field for the homogeneous equation \eqref{hmneq}  in terms of Fourier modes can be expressed as
\begin{equation}
    \label{hmnult}
    h_{\mu\nu}(x) = \int d^3\tilde{p} \sum_{\lambda = 1}^2[\varepsilon_{\mu\nu}^{(\lambda)} (p_{0+}, {\bf p}) a^{(\lambda)}(p_{0+}, {\bf p}) e^{-i p \cdot x} + \varepsilon_{\mu\nu}^{*(\lambda)} (p_{0+}, {\bf p}) a^{\dagger(\lambda)}(p_{0+}, {\bf p}) e^{i p \cdot x} ],
\end{equation}where $$d^3\tilde{p} = \frac{d^3p}{(2\pi)^32\left|{\bf p}\right|\sqrt{1 + \xi b_0^2 - \xi\left|{\bf b}\right|^2 \cos^2\Psi} }.$$

Now, using the Equation \eqref{eq:const6} with the traceless condition and choosing the direction of propagation in $z$ axis, the four-momentum is $p^\mu = (p_0, 0, 0, p_3)$. Therefore, we can see the modifications on the polarization tensor. We will consider some special cases.

First, we consider $b^\mu = (b_0, 0, 0, 0)$ or $b^\mu = (b_0, 0, 0, b_3)$. In this case, the polarization tensor is
\begin{equation}
\varepsilon^{\mu\nu} =
\left(
\begin{array}{cccc}
0 & 0 & 0 & 0 \\
0 & \varepsilon^{11} & \varepsilon^{12} & 0  \\
0 & \varepsilon^{12} & - \varepsilon^{11} & 0 \\
0 & 0 & 0 & 0 
\end{array}
\right).
\end{equation}

We see that this case is equal to the usual one. This is expected because in the first case, it was considered just a $b^\mu$ timelike and in the second one, no preferred direction was selected in the spacetime. However, if we consider the case where $b_\mu = (b_0, 0, b_2, 0)$, we have
\begin{equation}
\varepsilon^{\mu\nu} =
\left(
\begin{array}{cccc}
\varepsilon^{00} & \varepsilon^{10} & -\varepsilon^{00}\frac{b_0}{b_2} & -\varepsilon^{00}\frac{p_0}{p_3} \\
\varepsilon^{10} & -\varepsilon^{00}\left[1 + \frac{(b_0)^2}{(b_2)^2} + \frac{(p_0)^2}{(p_3)^2}\right] & -\varepsilon^{10} & - \varepsilon^{10}\frac{p_0}{p_3}  \\
-\varepsilon^{00}\frac{b_0}{b_2} & -\varepsilon^{10}\frac{b_0}{b_2} & \varepsilon^{00}\frac{(b_0)^2}{(b_2)^2} & \varepsilon^{00}\frac{b_0}{b_2}\frac{p_0}{p_3} \\
-\varepsilon^{00}\frac{p_0}{p_3} & -\varepsilon^{10}\frac{b_0}{b_2}\frac{p_0}{p_3} & \varepsilon^{00}\frac{b_0}{b_2}\frac{p_0}{p_3} & \varepsilon^{00}\frac{(p_0)^2}{(p_3)^2} 
\end{array}
\right).
\end{equation}

Hence, the existence of a vacuum expectation value of the bumblebee field yields to an LSB. We see that the polarization tensor components do have a dependence on components of the bumblebee field. So, the polarization tensor has a dependence on the direction of this vector field.

\section{Gravitational waves production in the presence of the bumblebee field}

For the analysis of the production of GW, we add a current $J_{\mu\nu}$ to the homogeneous equation to the graviton field in Eq. \eqref{hmneq} resulting in 
\begin{equation}
    [\Box + \xi (b \cdot \partial)^2]h_{\mu\nu}(x) = J_{\mu\nu}(x).
\end{equation}

The method of the Green function is commonly used to solve this kind of differential equation. The Green function must satisfy
\begin{equation}\label{eq:FGfunc}
    \Hat{\mathcal{O}} \; G(x-y) = \delta^{(4)}(x-y),
\end{equation}here $\Hat{\mathcal{O}} = \Box + \xi(b\cdot \partial)^2$. Finding $G(x-y)$, $h_{\mu\nu}$ is determined by the convolution integration
\begin{equation}
    h_{\mu\nu}(x) = \int d^4y \; G(x-y) J_{\mu\nu}(y).
\end{equation}

In the usual Einstein-Hilbert linearized theory, $G_R(x-y)$ is the retarded Green function
\begin{equation}\label{eq:retGf}
    G_{R}(x-y) = \frac{1}{4\pi r}\delta[\tau - r] \Theta(\tau),
\end{equation}here it was considered that the four-vectors are divided into a temporal and    a spatial part ($x^\mu = (x^0, {\bf x})$). In this scenario, $\tau = x^0 - y^0$, $r = |{\bf x} - {\bf y}|$ and $\Theta(x)$ is the Heaviside step function. The convolution leads to the quadrupole formula of $h_{\mu\nu}$ considering a nonrelativistic, isolated and far-away object.

For the modified graviton equation, we can represent the Green function in the momentum space, solving the Eq. (\ref{eq:FGfunc}) we get
\begin{equation}
    \Tilde{G}(p) = \frac{1}{p^2 + \xi (b \cdot p)^2},
\end{equation}so that to get $G$ in the configuration space, we may use the inverse Fourier transform. The inversion is made for two particular cases; the first is considering that $b^\mu$ has a timelike configuration $b^\mu = \left(b^0, 0\right)$, in this case, the Green function resumes to
\begin{equation}
    \Tilde{G}(p) = \frac{1}{[1+\xi b_{0}^{2}]p_0^2 - \left|{\bf p}\right|},
\end{equation}and the inversion results to
\begin{equation}
    G(x-y) = \frac{1}{4\pi \sqrt{1+\xi b_{0}^{2}} r}\delta\left[r - \frac{\tau}{\sqrt{1+\xi b_{0}^{2}}}\right].
\end{equation}

The second case is a spacelike configuration $b^\mu = (0,{\bf b})$, which will reduce $\Tilde{G}(p)$ to
\begin{equation}\label{eq:ftGbvec}
    \Tilde{G}(p) = \frac{1}{p_0^2 - (1-\xi |{\bf b}|^2 \cos^2{\Psi})|{\bf p}|},
\end{equation}where $\Psi$ is the angle between ${\bf p}$ and ${\bf b}$, therefore $\cos^2{\Psi} = \cos(\theta) \cos(\theta_b) + \sin(\theta) \sin(\theta_b) \cos(\phi_b - \phi)$, with ($\theta$, $\phi$) and ($\theta_b$, $\phi_b$) the angular coordinates of the trivectors ${\bf p}$ and ${\bf b}$, respectively.

The inverse Fourier transform of Eq. (\ref{eq:ftGbvec}) expanded to the second order of $|{\bf b}|$ is
\begin{equation}
    G(x-y) = G_R(x-y) - G_2(x-y),
\end{equation}with the second term being given by
\begin{equation}
    \begin{split}
        G_2(x-y) = & \frac{\xi}{8\pi r^3} \Theta(\tau) \times \\
        & \times \left\{(b \cdot r)^2 \left[ \left(\frac{\tau}{r} -1 \right)\delta(\tau - r) + \tau \delta'(\tau - r) \right] + \tau r b^2 \cos({2\theta_b}) \; \delta(\tau - r)\right\}.
    \end{split}
\end{equation}

For the usual Einstein-Hilbert linearized theory we have the formula for the perturbation
\begin{equation}
    h_{ij}(t,{\bf r}) = \frac{2G}{r}\frac{d^2 I_{ij}}{dt^2}(t_r),
\end{equation}where $t_r = t - r$ is the retarded time and $I_{ij}$ is the quadrupole momentum defined as
\begin{equation}
    I_{ij}(t) = \int y^i y^j T^{00} d^3y.
\end{equation}

For the modified equation with $b^\mu = (b^0, 0)$ and considering that $J_{\mu\nu} = 16\pi G T_{\mu\nu}$ the perturbation is
\begin{equation}
    h_{ij}(t,{\bf r}) = \frac{2G}{r}\frac{d^2 I_{ij}}{dt^2}(t'_r),
\end{equation}here, the retarded time is modified to $t'_r = t - r\sqrt{1+b_{0}^{2}}$, evidencing that the waves propagates slower.

When $b^\mu = (0, {\bf b})$ the perturbation is
\begin{equation}
     h_{ij}(t,{\bf r}) = \frac{2G}{r} \left[ \left( 1-\frac{\xi b^2}{2}\cos{2\theta_b}\right)\frac{d^2I_{ij}}{dt^2}(t_r) + \frac{\xi ({\bf b}\cdot {\bf r})^2}{2r} \frac{d^3 I_{ij}}{dt^3}(t_r)\right].
\end{equation}

In this equation, we can see the anisotropy in the solution; it is a well-known feature of theories modified by the bumblebee vector \cite{MalufGravity1}. The vector ${\bf b}$ selects a preferential direction for the wave propagation. Another peculiarity is the third derivative of the $I_{ij}$ that appeared in the solution.

\section{Conclusion}
In this work, we investigated the effects of the Lorentz symmetry breaking due to the presence of a non-zero VEV for a vector field. We use the simplest model in literature, the so-called bumblebee model. We expand the density Lagrangian of the theory up to the second order of $h_{\mu\nu}$, and we find the modified equation of motion to the graviton. Hence, we obtain the modified dispersion relation and the corresponding plane wave solution. We also show that the free graviton has only two degrees of freedom, and we explicitly determine the new polarization tensor dependent on the background field.

In the sequence, we seek for the effects of the spontaneous Lorentz breaking in the production of the gravitational waves. Assuming the presence of a matter source, we determine the retarded Green function and corresponding solution for the graviton field in two special configurations. In the first one, we considered a timelike configuration for $b^\mu$. In this case, only the velocity of propagation of the wave was smaller by a factor of $\sqrt{1+\xi b_{0}^{2}}$. In the second case, we considered a spacelike configuration for the bumblebee VEV. The corrections to the quadrupole formula showed the existence of anisotropy in the solution, showing an explicit dependency in the relative direction between the background field and the position vector.

\section{Acknowledgments}

\hspace{0.5cm}The authors would like to thank the Funda\c{c}\~{a}o Cearense de apoio ao Desenvolvimento Cient\'{\i}fico e Tecnol\'{o}gico (FUNCAP), the Coordena\c{c}\~ao de Aperfei\c{c}oamento de Pessoal de N\'ivel Superior (CAPES), and the Conselho Nacional de Desenvolvimento Cient\'{\i}fico e Tecnol\'{o}gico (CNPq) for
financial support. The work by RVM has been supported by the
CNPq project No. 305678/2015-9.


\end{document}